# Ferromagnetic films with three to twenty spin layers as described using second order perturbed Heisenberg Hamiltonian

P. Samarasekara and T.H.Y.I.K. de Silva

Department of Physics, University of Peradeniya, Peradeniya, Sri Lanka

**Abstract**

Modified second order perturbed Heisenberg Hamiltonian was employed to describe the magnetic properties of ferromagnetic films with three to twenty spin layers for the first time. Previously, the solution of second order perturbed Heisenberg Hamiltonian was found for ferromagnetic films with two to five layers under some assumptions by us. In this report, the exact solution is presented without any assumptions by calculating the pseudo inverse of matrix *C*. All eight magnetic parameters such as spin exchange interaction, second order magnetic anisotropy, fourth order magnetic anisotropy, stress induced magnetic anisotropy, demagnetization factor, in plane magnetic field, out of plane magnetic field and magnetic dipole interaction were taken into account. The easy and hard directions were found using 3-D and 2-D plots of total magnetic energy versus these magnetic parameters. Angle between easy and hard directions was nearly 107 degrees in many cases. The magnetic easy axis gradually rotates from out of plane to in plane direction by indicating a preferred in plane orientation of magnetic easy axis for films with higher number of spin layers. These theoretical results agree with some experimental research data of ferromagnetic thin films.

## 1. Introduction:

Ferromagnetic thin films find potential applications in magnetic memory and microwave devices due to their unique properties. In compact devices, ferromagnetic thin films replace the bulk ferromagnetic materials. However, a systematic detailed theoretical study of ferromagnetic ultra thin films is not available. Classical, semi classical and quantum theoretical models are employed to describe the ferromagnetic films. While the magnitude of spin continuously varies from 0 to 1 in classical models, the spin has some discrete values in quantum models.

Ferromagnetic thin films have been studied using the Heisenberg Hamiltonian with spin exchange interaction, magnetic dipole interaction, applied magnetic field, second and fourth order magnetic anisotropy terms [1, 2, 3]. Domain structure and Magnetization reversal in thin magnetic films have been theoretically investigated [4]. In-plane dipole coupling anisotropy of a square ferromagnetic Heisenberg monolayer has been explained using Heisenberg Hamiltonian [5]. Effect of the interfacial coupling on the magnetic ordering in ferro-antiferromagntic bilayers has been studied using the Heisenberg Hamiltonian [6].

In addition, EuTe films with surface elastic stresses have been theoretically studied using Heisenberg Hamiltonian [7]. Magnetostriction of dc magnetron sputtered FeTaN thin films has been theoretically studied using the theory of De Vries [8]. Magnetic layers of Ni on Cu have been theoretically investigated using the Korringa-Kohn-Rostoker Green's function method [9]. Electric and magnetic properties of multiferroic thin films have been theoretically explained by modified Heisenberg and transverse Ising model using Green's function technique [10]. The quasistatic magnetic hysteresis of ferromagnetic thin films grown on a vicinal substrate has been theoretically investigated by Monte Carlo simulations within a 2D model [11]. Structural and magnetic properties of two dimensional FeCo orders alloys deposited on W(110) substrates have been studied using first principles band structure theory [12].

Previously nickel ferrite films were synthesized using sputtering by us [13]. In addition, lithium mixed ferrite films were fabricated using pulsed laser deposition [14]. For all these films, the coercivity of film increased due to the stress induced anisotropy. The change of coercivity due to the stress induced anisotropy was qualitatively calculated for all these films. The calculated values of the change of coercivity agreed with the experimentally found values. So the stress induced anisotropy plays a major role in magnetic thin fabrications. Previously the Heisenberg Hamiltonian was employed to investigate the second order perturbed energy of ultrathin ferromagnetic films [15], unperturbed energy of thick ferromagnetic films [16], third order perturbed energy of thick spinel ferrite [17], third order perturbed energy of thin spinel ferrite [18], second order perturbed energy of thick spinel ferrite films [19], spin reorientation of barium ferrite [20], third order perturbed energy of ultra-thin ferromagnetic films [21], second order perturbed energy of ferrite thin films [22] and spin reorientation of nickel ferrite films [23]. The magnetic dipole interaction and demagnetization factor are microscopic and macroscopic effects, respectively. Therefore, both these terms were taken into consideration in our model.

## 2. Model:

The Heisenberg Hamiltonian of ferromagnetic films is given as following.

$$H = -\frac{J}{2}\sum_{m,n}\vec{S}_m.\vec{S}_n + \frac{\omega}{2}\sum_{m\neq n}(\frac{\vec{S}_m.\vec{S}_n}{r_{mn}^{3}} - \frac{3(\vec{S}_m.\vec{r}_{mn})(\vec{r}_{mn}.\vec{S}_n)}{r_{mn}^{5}}) - \sum_m D_{\lambda_m}^{(2)}(S_m^z)^2 - \sum_m D_{\lambda_m}^{(4)}(S_m^z)^4$$

$$-\sum_{m,n}[\vec{H} - (N_d\vec{S}_n/\mu_0)].\vec{S}_m - \sum_m K_s Sin2\theta_m$$

Here $\vec{S}_m$ and $\vec{S}_n$ are two spins. Above equation can be deduced to following form.

$$E(\theta) = -\frac{1}{2}\sum_{m,n=1}^{N}[(JZ_{|m-n|} - \frac{\omega}{4}\Phi_{|m-n|})\cos(\theta_m - \theta_n) - \frac{3\omega}{4}\Phi_{|m-n|}\cos(\theta_m + \theta_n)]$$

$$-\sum_{m=1}^{N}(D_m^{(2)}\cos^2\theta_m + D_m^{(4)}\cos^4\theta_m + H_{in}\sin\theta_m + H_{out}\cos\theta_m)$$

$$+\sum_{m,n=1}^{N}\frac{N_d}{\mu_0}\cos(\theta_m - \theta_n) - K_s\sum_{m=1}^{N}\sin 2\theta_m \qquad (1)$$

Where $J, Z_{|m-n|}$, $\omega$, $\Phi_{|m-n|}$, $\theta$, $D_m^{(2)}, D_m^{(4)}, H_{in}, H_{out}, N_d, K_s$, $m$, $n$ and $N$ are spin exchange interaction, number of nearest spin neighbors, strength of long range dipole interaction, constants for partial summation of dipole interaction, azimuthal angle of spin, second and fourth order anisotropy constants, in plane and out of plane applied magnetic fields, demagnetization factor, stress induced anisotropy constant, spin plane indices and total number of layers in film, respectively.

In terms of small perturbations $\varepsilon_m$ and $\varepsilon_n$, the azimuthal angles of spins can be expressed as $\theta_m = \theta + \varepsilon_m$ and $\theta_n = \theta + \varepsilon_n$. After substituting these new angles in above equation number 1, the cosine and sine terms can be expanded up to the second order of $\varepsilon_m$ and $\varepsilon_n$ as following.

$E(\theta) = E_0 + E(\varepsilon) + E(\varepsilon^2)$ **(2)**

Here $E_0 = -\dfrac{1}{2}\sum_{m,n=1}^{N}(JZ_{|m-n|} - \dfrac{\omega}{4}\Phi_{|m-n|}) + \dfrac{3\omega}{8}\cos 2\theta \sum_{m,n=1}^{N}\Phi_{|m-n|}$

$$-\cos^2\theta \sum_{m=1}^{N}D_m^{(2)} - \cos^4\theta \sum_{m=1}^{N}D_m^{(4)} - N(H_{in}\sin\theta + H_{out}\cos\theta - \dfrac{N_d}{\mu_0} + K_s\sin 2\theta)\ \textbf{(3)}$$

$E(\varepsilon) = -\dfrac{3\omega}{8}\sin 2\theta \sum_{m,n=1}^{N}\Phi_{|m-n|}(\varepsilon_m + \varepsilon_n) + \sin 2\theta \sum_{m=1}^{N}D_m^{(2)}\varepsilon_m + 2\cos^2\theta\sin 2\theta\sum_{m=1}^{N}D_m^{(4)}\varepsilon_m$

$$-H_{in}\cos\theta\sum_{m=1}^{N}\varepsilon_m + H_{out}\sin\theta\sum_{m=1}^{N}\varepsilon_m - 2K_s\cos 2\theta\sum_{m=1}^{N}\varepsilon_m$$

$E(\varepsilon^2) = \dfrac{1}{4}\sum_{m,n=1}^{N}(JZ_{|m-n|} - \dfrac{\omega}{4}\Phi_{|m-n|})(\varepsilon_m - \varepsilon_n)^2 - \dfrac{3\omega}{16}\cos 2\theta\sum_{m,n=1}^{N}\Phi_{|m-n|}(\varepsilon_m + \varepsilon_n)^2$

$$-(\sin^2\theta - \cos^2\theta)\sum_{m=1}^{N}D_m^{(2)}\varepsilon_m^2 + 2\cos^2\theta(\cos^2\theta - 3\sin^2\theta)\sum_{m=1}^{N}D_m^{(4)}\varepsilon_m^2$$

$$+\dfrac{H_{in}}{2}\sin\theta\sum_{m=1}^{N}\varepsilon_m^2 + \dfrac{H_{out}}{2}\cos\theta\sum_{m=1}^{N}\varepsilon_m^2 - \dfrac{N_d}{2\mu_0}\sum_{m,n=1}^{N}(\varepsilon_m - \varepsilon_n)^2$$

$$+2K_s\sin 2\theta\sum_{m=1}^{N}\varepsilon_m^2$$

After using the constraint $\sum_{m=1}^{N}\varepsilon_m = 0$ , $E(\ \varepsilon) = \vec{\alpha}.\vec{\varepsilon}$

Here $\vec{\alpha}(\varepsilon) = \vec{B}(\theta)\sin 2\theta$ are the terms of matrices with

$$B_\lambda(\theta) = -\dfrac{3\omega}{4}\sum_{m=1}^{N}\Phi_{|\lambda-m|} + D_\lambda^{(2)} + 2D_\lambda^{(4)}\cos^2\theta \quad \textbf{(4)}$$

Also $E(\varepsilon^2) = \dfrac{1}{2}\vec{\varepsilon}.C.\vec{\varepsilon}$

Here the elements of matrix are given by,

$$C_{mn} = -(JZ_{|m-n|} - \frac{\omega}{4}\Phi_{|m-n|}) - \frac{3\omega}{4}\cos 2\theta\Phi_{|m-n|} + \frac{2N_d}{\mu_0}$$

$$+ \delta_{mn}\{\sum_{\lambda=1}^{N}[JZ_{|m-\lambda|} - \Phi_{|m-\lambda|}(\frac{\omega}{4} + \frac{3\omega}{4}\cos 2\theta)] - 2(\sin^2\theta - \cos^2\theta)D_m^{(2)}$$

$$+ 4\cos^2\theta(\cos^2\theta - 3\sin^2\theta)D_m^{(4)} + H_{in}\sin\theta + H_{out}\cos\theta - \frac{4N_d}{\mu_0} + 4K_s\sin 2\theta\} \quad \textbf{(5)}$$

Therefore, the total magnetic energy given in equation 2 can be deduced to

$$E(\theta)=E_0 + \vec{\alpha}.\vec{\varepsilon} + \frac{1}{2}\vec{\varepsilon}.C.\vec{\varepsilon} = E_0 - \frac{1}{2}\vec{\alpha}.C^+.\vec{\alpha} \qquad \textbf{(6)}$$

Here $C^+$ is the pseudo-inverse given by

$$C.C^+ = 1 - \frac{E}{N}. \qquad \textbf{(7)}$$

Here $E$ is the matrix with all elements $E_{mn}=1$.

## 3. Results and Discussion:

First matrix elements of $C$ were found using equation 5. Then elements of $C^+$ were determined using equation 7 and MATLAB computer program. Finally total energy was found using equation 6. When one magnetic parameter was varied, other parameters were kept at constant values. Figure 1 shows the 3-D plot of $\frac{E(\theta)}{\omega}$ versus $\frac{D_m^{(4)}}{\omega}$ and angle for ferromagnetic thin films with simple cubic (s. c.) lattice. All the graphs in figure 1 to figure 4 are given for ferromagnetic films with 3 spin layers ($N$=3). Other parameters were kept at constant values as following.

$$\frac{J}{\omega} = \frac{D_m^{(2)}}{\omega} = \frac{K_s}{\omega} = \frac{N_d}{\mu_0\omega} = \frac{H_{in}}{\omega} = \frac{H_{out}}{\omega} = 10$$

For s. c. (001) lattice, $Z_0=4$, $Z_1=1$, $\Phi_0=9.0336$ and $\Phi_1=-0.3275$ [1, 2, 3].

Several energy maximums and minimums can be observed in this 3-D plot. One energy minimum and maximum of this 3-D plot can be found at $\dfrac{D_m^{(4)}}{\omega}=35$ and 25, respectively.

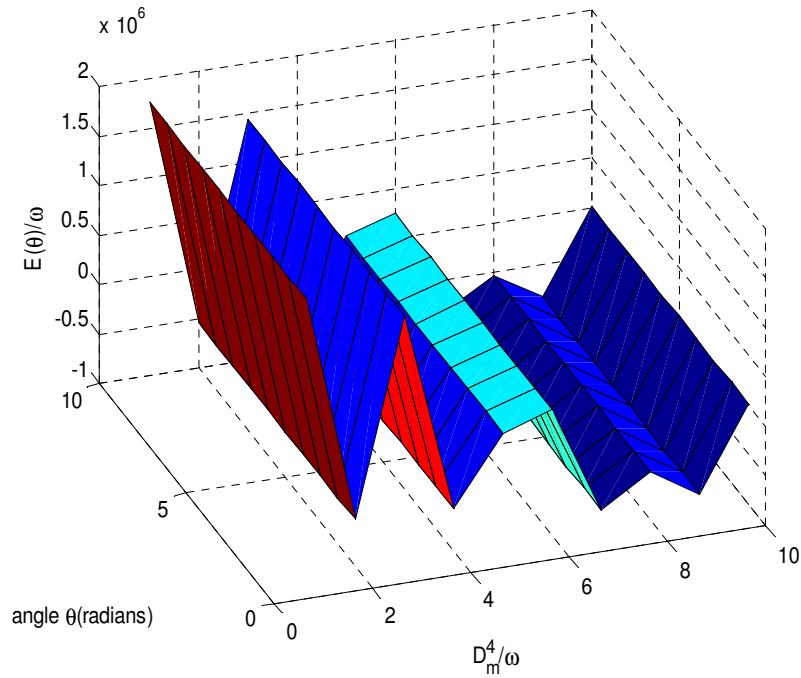

Figure 1: 3-D plot of $\dfrac{E(\theta)}{\omega}$ versus $\dfrac{D_m^{(4)}}{\omega}$ and angle for s. c. lattice

Figure 2 shows the graph of $\dfrac{E(\theta)}{\omega}$ versus angle at $\dfrac{D_m^{(4)}}{\omega}=35$. A minimum and a maximum of this plot can be observed at 10°54′ and 123°28′, respectively. Minimum and maximum of energy correspond to the magnetic easy and hard directions.

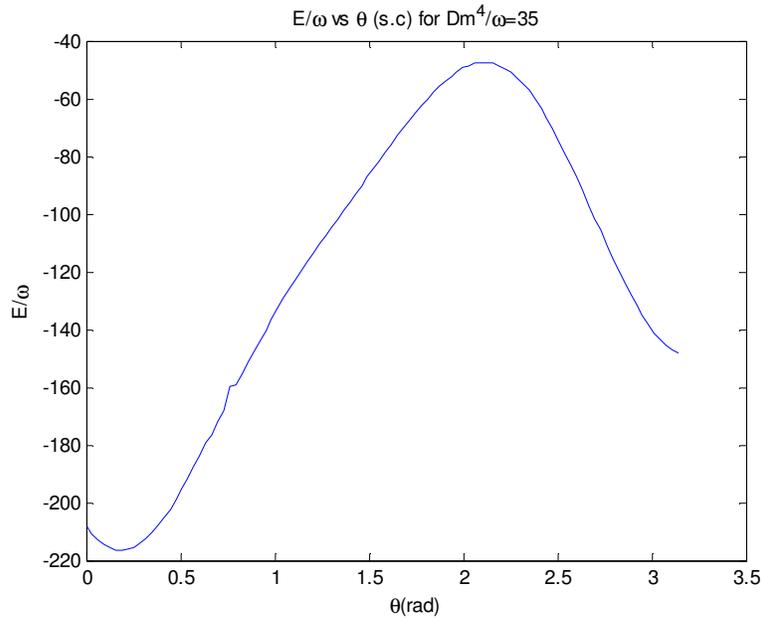

Figure 2: Graph of $\frac{E(\theta)}{\omega}$ versus angle at $\frac{D_m^{(4)}}{\omega}$ =35.

Figure 3 shows the graph of $\frac{E(\theta)}{\omega}$ versus angle at $\frac{D_m^{(4)}}{\omega}$ =25. A minimum and a maximum of this plot can be observed at 12 °42 ′ and 123 °38 ′, respectively. According to Figure 2 and 3, real easy and hard directions can be observed at 10°54′ and 123 °38 ′, respectively.

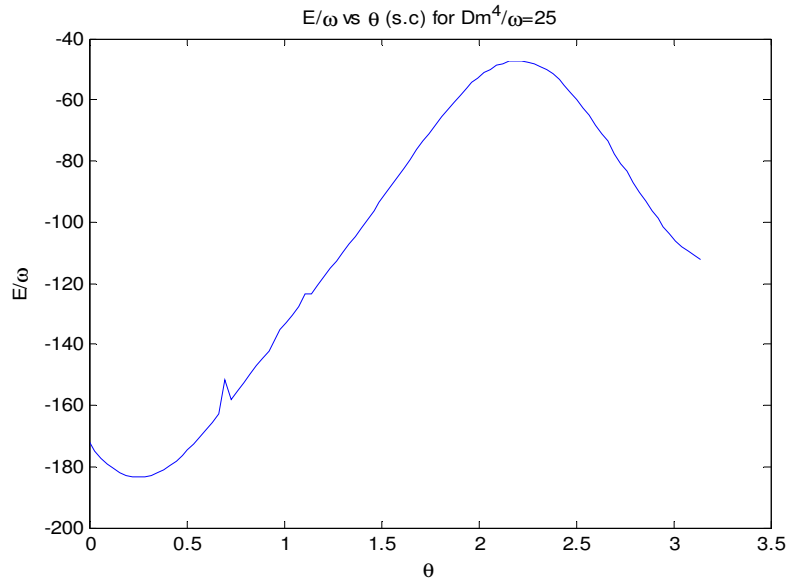

Figure 3: Graph of $\frac{E(\theta)}{\omega}$ versus angle at $\frac{D_m^{(4)}}{\omega} = 25$

Figure 4 shows the 3-D plot of $\frac{E(\theta)}{\omega}$ versus angle and $\frac{D_m^{(2)}}{\omega}$. Other magnetic parameters were kept at $\frac{J}{\omega} = \frac{D_m^{(4)}}{\omega} = \frac{K_s}{\omega} = \frac{N_d}{\mu_o \omega} = \frac{H_{in}}{\omega} = \frac{H_{out}}{\omega} = 10$. Several energy maximums and minimums appear in this 3-D plot. The shape of this 3-D plot is different from the 3-D plot given in figure 1. One energy minimum and maximum can be observed at $\frac{D_m^{(2)}}{\omega} = 58$ and 65, respectively. The real easy direction was found to be 10°54′ from the graph of $\frac{E(\theta)}{\omega}$ versus angle plotted for $\frac{D_m^{(2)}}{\omega} = 58$. Similarly, the real hard direction was found to be 103°38′ from the graph of $\frac{E(\theta)}{\omega}$ versus angle plotted for $\frac{D_m^{(2)}}{\omega} = 65$. All the data in this manuscript are given for s. c. structure.

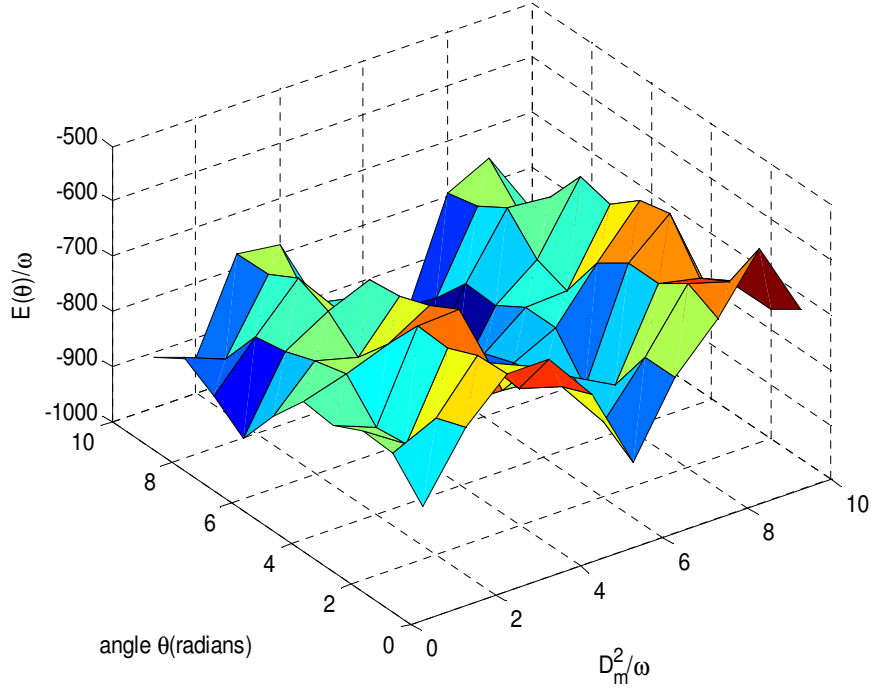

Figure 4: 3-D plot of $\dfrac{E(\theta)}{\omega}$ versus angle and $\dfrac{D_m^{(2)}}{\omega}$.

Similarly easy and hard directions were calculated for ferromagnetic films with s. c. structure using plots of $\dfrac{E(\theta)}{\omega}$ versus angle for $\dfrac{J}{\omega} = \dfrac{D_m^{(4)}}{\omega} = \dfrac{K_s}{\omega} = \dfrac{N_d}{\mu_o \omega} = \dfrac{H_{in}}{\omega} = \dfrac{H_{out}}{\omega} = \dfrac{D_m^{(2)}}{\omega} 10$ and each value of number of layers. These data are given in table 1. Columns 3, 5, 6 and 7 represent the $\dfrac{E(\theta)}{\omega}$ of easy direction, $\dfrac{E(\theta)}{\omega}$ of hard direction, energy difference between easy and hard directions and angle between easy and hard directions, respectively. Column 6 shows the energy required to rotate spins or magnetic moments from easy direction to hard direction ($\Delta$E). In applications, materials with less $\Delta$E are preferred. The energy required to rotate spins or magnetic moments from easy direction to hard direction gradually increases with number of layers. The angles between easy and hard directions are close to 107 degrees. Here $\theta$ is the angle between a normal drawn to film plane and the spin. Magnetic easy direction gradually rotates from out of plane to in plane direction with increase of number of layers.

These data agree with the experimental date obtained for some ferromagnetic thin films. The magnetic easy axis of sputter synthesized ferromagnetic Ni films with lattice parameter 0.352 nm indicates a preferred in plane orientation at higher thicknesses [24]. According to this experimental data, the spin reorientation transition occurs from in plane to out plane at film thickness in between 14 and 24 $^0$A. The magnetic easy axis of ferromagnetic Fe thin films deposited by electron beam evaporation also rotates from out of plane to in plane direction at the film thickness of 2 monolayers, as the film thickness is increased [25]. Deposition temperature, annealing temperature, orientation of substrate, type of sputtering gas, sputtering pressure, deposition rate and film thickness govern the orientation of easy axis of thin films [13, 14]. Previously, the variation of easy axis orientation of magnetic thin films with temperature has been explained by us [20, 23]. Magnetic energy due to spin exchange interaction, second order magnetic anisotropy, fourth order magnetic anisotropy, magnetic field, stress induced anisotropy decreases with the number of layers. On the other hand, magnetic energy due to magnetic dipole interaction and demagnetization factor increases with number of layers. The reason for in plane orientation of easy axis is attributed to the domination of energy due to magnetic dipole interaction and demagnetization factor at higher thicknesses.

| N (Number of spin layers) | θ(easy) In degrees | E/ω (easy) | θ(hard) In degrees | E/ω (hard) | ΔE= E(easy)-E(hard) | Δθ=θ(hard) -θ(easy) In degrees |
|---|---|---|---|---|---|---|
| 3 | 27.1237 | -148.4703 | 136.3637 | -39.2801 | 109.1902 | 109.2431 |
| 4 | 27.2727 | -201.3010 | 136.3638 | -55.7303 | 145.5711 | 109.0911 |
| 5 | 32.72735 | -243.2938 | 138.1802 | -65.5296 | 177.7642 | 105.4529 |
| 6 | 32.72735 | -293.0313 | 140.0022 | -80.3742 | 212.6571 | 107.2749 |
| 7 | 32.72735 | -342.4569 | 140.0022 | -94. 9925 | 247.4644 | 107.2749 |
| 8 | 34.54363 | -391.6847 | 141.8185 | -109.4450 | 282.2397 | 107.2749 |
| 9 | 34.54363 | -440.6979 | 141.8185 | -123.5873 | 317.1106 | 107.2749 |
| 10 | 36.36563 | -489. 9321 | 143.6348 | -136. 9551 | 352.977 | 107.2692 |
| 11 | 36.36563 | -538.2348 | 143.6348 | -151.1074 | 387.1274 | 107.2692 |
| 12 | 38.18191 | -586.7906 | 145.4568 | -164.3518 | 422.4388 | 107.2749 |
| 13 | 38.18191 | -635.3465 | 145.4568 | -177.2214 | 458.1251 | 107.2749 |
| 14 | 39.99818 | -683.8709 | 147.2731 | -189.5887 | 494.2822 | 107.2749 |
| 15 | 39.99818 | -732.5906 | 147.2731 | -201.3705 | 531.2201 | 107.2749 |
| 16 | 41.82019 | -781.2445 | 149.0893 | -212.7722 | 568.4723 | 107.2692 |
| 17 | 41.82019 | -829.7926 | 149.0893 | -233. 9804 | 595.8122 | 107.2692 |
| 18 | 43.63647 | -878.5634 | 150.9114 | -234.6233 | 643.9401 | 107.2749 |
| 19 | 43.63647 | -927.5520 | 156.3659 | -243.7946 | 683.7574 | 112.7294 |
| 20 | 45.45274 | -977.3222 | 152.7276 | -253.1361 | 724.1861 | 107.2749 |

Table 1: Magnetic easy and hard directions for s. c. structured ferromagnetic films

## 4. Conclusion:

The magnetic properties of s. c. structured ferromagnetic films were investigated using modified second order Heisenberg Hamiltonian. Several easy and hard directions could be observed in many 3-D plots. Angle between easy and hard directions was close to 107 degrees in this range of thickness. The energy required to rotate spins from easy to hard direction in ferromagnetic films gradually increases with the number of spin layers. The magnetic easy direction of ferromagnetic films gradually rotates from the perpendicular direction to the in plane direction, as the number of spin layers is increased. This implies that a preferred in plane orientation can be observed at higher thicknesses. Our theoretical results agree with the experimental data of ferromagnetic thin films. Although the data of easy and hard directions given table 1 were calculated using graphs of $\dfrac{E(\theta)}{\omega}$ versus angle for $\dfrac{J}{\omega} = \dfrac{D_m^{(4)}}{\omega} = \dfrac{K_s}{\omega} = \dfrac{N_d}{\mu_o \omega} = \dfrac{H_{in}}{\omega} = \dfrac{H_{out}}{\omega} = \dfrac{D_m^{(2)}}{\omega} 10$, the same simulation can be carried out for other values of these magnetic energy parameters as well. The easy axis rotates from out of plane to in plane direction with the increase of thickness of ferromagnetic film, because demagnetization factor and dipole interaction dominates the magnetic anisotropies with the increase of thickness.